\documentclass[referee,a4paper,12pt,traditabstract]{jswsc} 

\usepackage{graphicx}
\usepackage{subfigure}
\usepackage{epstopdf}
\usepackage[mathlines]{lineno}
\usepackage[authoryear,round]{natbib}
\usepackage[backref]{hyperref}
\usepackage{url}
\usepackage{aas_macros}
\usepackage{xspace}
\usepackage[version=3]{mhchem}
\usepackage{txfonts}
\newcommand{\mass}{\ensuremath{{\cal{M}}}}

\hypersetup{colorlinks=true,citecolor=cyan,urlcolor=cyan,linkcolor=blue}

\newcommand{\PSUAA}{Department of Astronomy and Astrophysics, 525 Davey Laboratory, 251 Pollock Road, Penn State, University Park, PA, 16802, USA}
\newcommand{\PSUCEHW}{Center for Exoplanets and Habitable Worlds, 525 Davey Laboratory, 251 Pollock Road, Penn State, University Park, PA, 16802, USA}
\newcommand{\PSETI}{Penn State Extraterrestrial Intelligence Center, 525 Davey Laboratory, 251 Pollock Road, Penn State, University Park, PA, 16802, USA}

\newcommand{\Ro}{\ensuremath{\mathrm{Ro}}\xspace}
\newcommand{\CaHK}{\ce{Ca} {\sc ii} H \&\ K\xspace}
\newcommand{\eg}{e.g.\@\xspace}
\newcommand{\ie}{i.e.\@\xspace}
\newcommand{\rphk}{\ensuremath{R^\prime_\mathrm{HK}}\xspace}

\newcommand{\about}{\ensuremath{{\sim}}}

\begin{document}


   \title{The ``Other'' Centuries-Long Record of Solar Magnetic Activity Cycles: Lessons From the Stars}

   
   \titlerunning{Stellar Activity Cycles}

   \authorrunning{Wright, Metcalfe, Petrovay, and Hayakawa}

   \author{Jason T. Wright
          \inst{1,}
          \inst{2,}
          \inst{3}
          \and
          Travis S. Metcalfe
          \inst{4}
        \and
          Kristof Petrovay
          \inst{5}
      \and
          Hisashi Hayakawa
          \inst{6,}
          \inst{7,}
          \inst{8,}
          \inst{9}          
        }

   \institute{\PSUAA \phantom{ } \email{\href{mailto:astrowright@gmail.com}{astrowright@gmail.com}}
         \and
             \PSUCEHW
        \and    
            \PSETI
        \and
             Center for Solar-Stellar Connections, WDRC, 9020 Brumm Trail, Golden, CO 80403, USA
        \and
            ELTE E\"otv\"os Lor\'and University, Department of Astronomy, Budapest,          Hungary        \and
            Institute for Space-Earth Environmental Research, Nagoya University, Nagoya, Japan \phantom{  }\email{\href{mailto:hisashi@nagoya-u.jp}{hisashi@nagoya-u.jp}}
        \and
            Institute for Advanced Research, Nagoya University, Nagoya, Japan
        \and
            Space Physics and Operations Division, RAL Space, Science and Technology Facilities Council, Rutherford Appleton Laboratory, Harwell Oxford, Didcot, UK
        \and
            Astro-Glaciology Laboratory, Riken Nishina Centre, Wako, Japan
    }


 
  \abstract
{
  Studies of space climate and solar magnetic activity and studies of stellar dynamos and magnetic cycles are separated by discipline and by the quality of the data we have to work with: the Sun gives us detailed knowledge of a single star, while the stars give us comparatively sparse data across a range of (sometimes poorly known) masses and ages.  This review, a collaboration among scientists with expertise in stellar astronomy, space climate, and solar physics, seeks to  provide a guide to what the Sun's and stars' magnetic activity records can tell us about how stellar dynamos operate on decadal, millennial, and stellar-evolutionary timescales. It reviewes the historical Solar sunspot record, terrestrial cosmogenic isotope records, and decades-long chromospheric activity measurements of dozens of sun-like stars across a range of masses, ages, and metallicities near the solar value. Particular emphasis is placed on the Maunder minimim, and the hunt for similar events in the stellar record, especially the recently discovered grand minimum event ongoing in the star HD 166620. These records show that solar-like dynamos are a common feature of Sun-like stars, and magnetic grand minimum events akin to the Maunder minimum might be very rare, with only one identified in 3,000 star-years of magnetic activity records. Together, solar and stellar dynamo studies are bringing a still-fuzzy picture into focus in which stellar magnetic activity weakens as stars spin down over billions of years towards a critical Rossby number near the solar value, at which point stellar cycles begin to be frustrated.  These periods of frustration manifest as magnetic grand minima, and increase in frequency until they become permanent and stars enter a period of very low and constant magnetic activity.
 }  


   \maketitle

\section{Scope and Purpose}

The Sun is a star, but in most ways we know far more about the Sun than any other star.  We can easily resolve the solar surface, we have more solar photons to study than most telescopes can handle, and we have instruments on Earth and throughout the solar system that can sample its extended atmosphere, wind, and magnetic fields.  

By contrast, we rarely know any global property of a star, even among the nearest ones, to better than 1\%, including its mass, radius, composition, rotation period, or luminosity.  Properties like a star's age, surface magnetic field strength, or wind flux are often uncertain to a factor of 2 or more. Because we can rarely resolve stellar surfaces, even basic features like starspot covering fraction, limb darkening, and differential rotation are very challenging to measure and so are unknown for most nearby stars. That said, Zeeman-Doppler imaging can provide magnetic field geometry for some stars \citep[see][]{Jeffers2023}, and in the last 30 years asteroseismology has become possible \citep[\eg][]{Kjeldsen1995,Kjeldsen1995b,Bedding04,Aerts2021}.

The different detail we can observe the Sun and stars have led to wide divisions in their studies. Heliophysicists, space physicists, space climatologists, and others focusing on the structure and impact of the Sun generally work far from the stellar astronomers both figuratively and literally, using different methods, attending different conferences, and publishing in different journals.

Stars do have one major advantage over the Sun however: as an ensemble, they allow us to observe how the Sun's structure and behavior depends on its age, mass, and composition (for instance, how stellar spindown proceeds for stars of different masses on metallicity, \eg \citealt{Curtis2020,Amard2020,Carvalho-Silva2025AgeActivity}).  The Sun and stars thus provide points of calibration for our models of stellar structure and evolution, and those models in turn inform us about aspects of the Sun and stars we cannot observe \citep[\eg][and references therein]{Choi2016}.

The stars can also inform us about properties of the Sun and Sun-like stars for which we do not yet have good models, such as their dynamos and magnetic activity variations, which is our focus here.  This is important because dynamos produce space weather and space climate that can have dramatic effects on the Earth, especially on our technology. In parallel, stellar astrophysicists are concerned with how the magnetic fields of stars affect the habitability of the exoplanets that orbit those stars \citep[\eg][]{Kay2019,Amaral2022,Gallet2017}, including how it might affect any technological species on those planets \citep[\eg][]{Cirkovic20}.

The ultimate aspiration of this review is to better connect the space climate and stellar astrophysics communities in their mutual quest for a better understanding of the long-term nature of stellar magnetic activity and its effect on planets. It begins with a review of the historical sunspot record and the Maunder Minimum, a discussion of the cosmogenic isotope proxy record and what it says about solar magnetic grand minima, and a brief review of the relevant aspects of solar dynamo theory. The second part of the paper serves primarily to introduce space climatologists to some of what we know about stellar magnetic activity cycles among cool stars like the Sun across time and mass. This paper thus seeks to be neither comprehensive nor novel, but to give a brief overview of how stellar astrophysicists think about solar magnetic activity in the context of stellar astrophysics (and vice versa) and to highlight a few key issues where connections between the disciplines might be most immediately fruitful. It concludes by offering an emerging, largely empirical model of stellar activity variations. 

\section{The Sunspot Cycle and Maunder Minimum}

\subsection{The Discovery of the Sunspot Cycle and the Maunder Minimum}

Humanity's first hints that the Sun was anything other than a diurnal and steady source of light and warmth above the clouds were the solar eclipses and sunspots. Such celestial phenomena have been sporadically reported for centuries by naked-eye observers since antiquity, especially in East Asia. Eclipse records with good calendar dates go back to the 8th century BCE at least \citep{wang1980,Stephenson1997,Hayakawa2025_eclipse}, and sunspot records back to the 2nd century BCE at least \citep{Wittmann1987,Yau1988,Willis2001,willis2018,Vaquero2009,Hayakawa2019_1769}. The existence and behavior of sunspots were swiftly documented and disseminated when instrumental solar observations started in Europe---since 1607 by camerae obscurae and since 1610 by telescopes \citep{Vaquero2009,Arlt2020,vokhmyanin2020,Hayakawa2024_Kepler}. Since then, sunspot observations continuing into modern times have constituted one of the longest running scientific experiments in human history \citep{Vaquero2009,Owens2013,Clette2023}. These records form a basis for construction and calibration of sunspot number series and sunspot group number series, quantifying solar magnetic activity \citep{Vaquero2009,munoz2019,Arlt2020,Clette2023,usoskin2023}.

These records distinctly show cyclic periodicity of sunspot number every $\approx$ 11.0 years, while their exact duration varies from 9.0 years (Solar Cycle 2) to 13.6 years (Solar Cycle 4) according to the International Sunspot Number Version 2 \citep{clette2016,Clette2023}. These periodicities are known as Schwabe Cycles after Heinrich Schwabe who is often credited with discovering this feature during decades of his sunspot observations \citep{Schwabe1843,hathaway2015}, although this feature had been already noticed at least by the mid 18th century elsewhere, by Horrebow, for example \citep{Hoyt1995,hathaway2015,Jorgensen2019}. Regardless, this means that it took more than a century of sunspot observations after Kepler and Harriot for astronomers to start explicitly addressing these solar cycles. The reason for this may have little to do with astronomers' diligence and much to do with the Sun itself: for approximately 70 years, from 1645--1715, there were only stray sunspot groups to be seen \citep{Usoskin2015,Vaquero2015,hayakawa2021_eimmart,hayakawa2024_early_MM}.

By the 1880s Sp\"orer and Maunder had documented the cycling activity robustly via daily sunspot counts \citep{Spoerer1890,Maunder1922}, and recognized the extreme dearth of sunspots in the record prior to 1715, a period dubbed the ``Maunder Minimum'' by \citet{Eddy76} (Fig.~\ref{fig:MM_SN_for_Jason_v2}). Though some have attributed the lack of sunspots in the record in this period to the record keepers and not the Sun, in a series of papers \citep[including][]{Eddy76,Eddy1983}, Jack Eddy mustered a variety of historical accounts, including observations of aurorae and the solar corona during eclipse, to (re-)establish the Maunder Minimum as a firm historical event.  


\subsection{Modern Understandings from Instrumental Records}

Today we understand that the Sun undergoes a roughly-11-year cycle of activity during which its surface magnetic field reorganizes, strengthens, then flips polarity and weakens \citep{hathaway2015,petrovay2020}. Such cycles have been instrumentally confirmed from 1607 onward, while their variability is under debate during the Maunder Minimum \citep{Vaquero2009,Arlt2020,Clette2023,Hayakawa2024_Kepler}. Strong fields on the surface of the Sun interact with the plasma of the photosphere, where they can produce sunspots, and the optically thin and rare plasma of the chromosphere, where they deposit energy that heats the gas, which can only cool via line emission, for instance via the resonance doublet of ionized calcium \citep[\eg][]{charbonneau2020,petrovay2020}.

\begin{figure}
    \centering
    \includegraphics[width=1\linewidth]{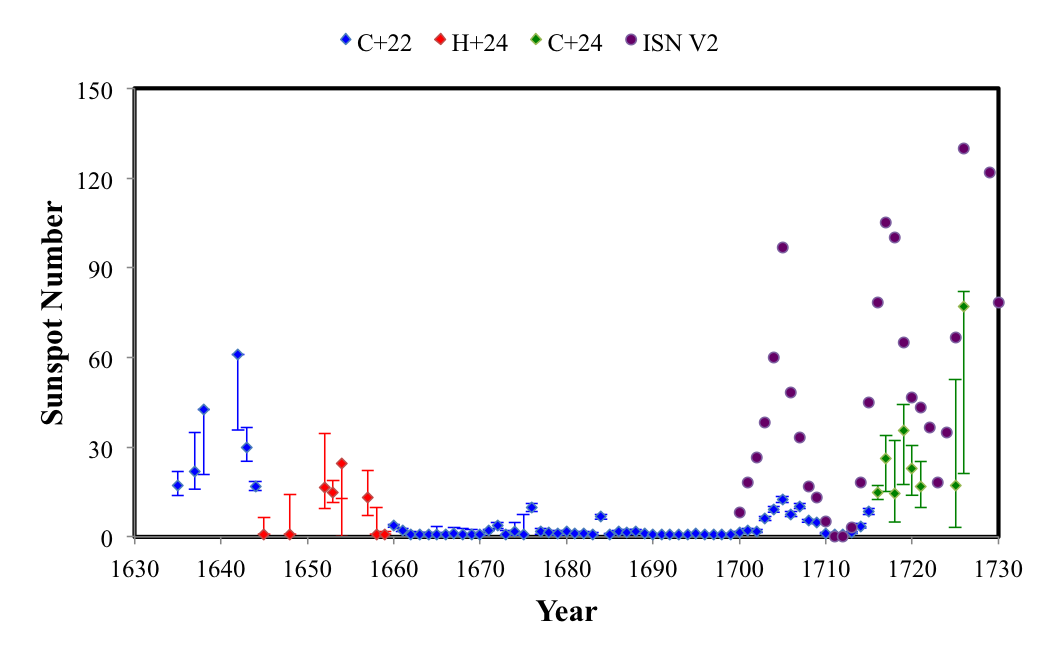}
    \caption{Sunspot number estimates for the Maunder Minimum based on the active-fraction methods (C+22 \citep{Carrasco2022}, H+24 \citep{hayakawa2024_early_MM}, and C+24 \citep{Carrasco2024}), in contrast with the standard yearly value of the International Sunspot Number Version 2 (ISN V2 \citep{clette2016}). H+24 superseded C+22 about their reconstructions in 1645-1659. Their estimates are to be revised slightly upward on the basis of the updated active day fractions \citep{hayakawa2025_review}. Some of the data points do not have an upper error margin. This is because the ADF saturates at 100 per cent as the solar activity level (ISN V2 in this case) goes up. ISN V2 does not have an error margin within this interval. }     
    \label{fig:MM_SN_for_Jason_v2}
\end{figure}

Both sunspot numbers and chromospheric emission are thus proxies for the overall field strength of the Sun, and track this field's evolution and cycles, and therefore the mechanism of the underlying solar dynamo. The Maunder Minimum, then, appears to represent some sort of substantial, but temporary, change in this dynamo's behavior \citep{Pevtsov2014,Egeland2017,Usoskin2021,usoskin2023}.

At least, this is what the scientific community has learned about the Maunder Minimum using multiple historical records and proxy data, building upon Eddy's benchmark studies \citep{Usoskin2015}. While some have argued that the depth and extent of the Maunder Minimum has been misinterpreted or overstated \citep{Zolotova2015}, extensive analysis of records of raw sunspot group counts and active day fraction during the Maunder Minimum have shown the Maunder Minimum (Fig.~\ref{fig:MM_SN_for_Jason_v2}) to be superlative with respect to other extended minima, such as the Dalton Minimum in the early 19th century, in terms of its solar cycle amplitude \citep{Kovaltsov2004,Vaquero2015,Usoskin2015,munoz2019,Carrasco2022,Carrasco2024,hayakawa2024_early_MM,hayakawa2025_review}. The Maunder Minimum also stands out in terms of the extreme hemispheric asymmetry \citep{ribes1993,hayakawa2021_eimmart}, the loss of apparent solar coronal structure \citep{Riley2015,hayakawa2021_eclipses}, and significant enhancement of production of cosmogenic isotopes \citep{beer1990,Muscheler2007,Usoskin2015,Muscheler2016,brehm2021}. 

\begin{figure}
    \centering
    \includegraphics[width=1\linewidth]{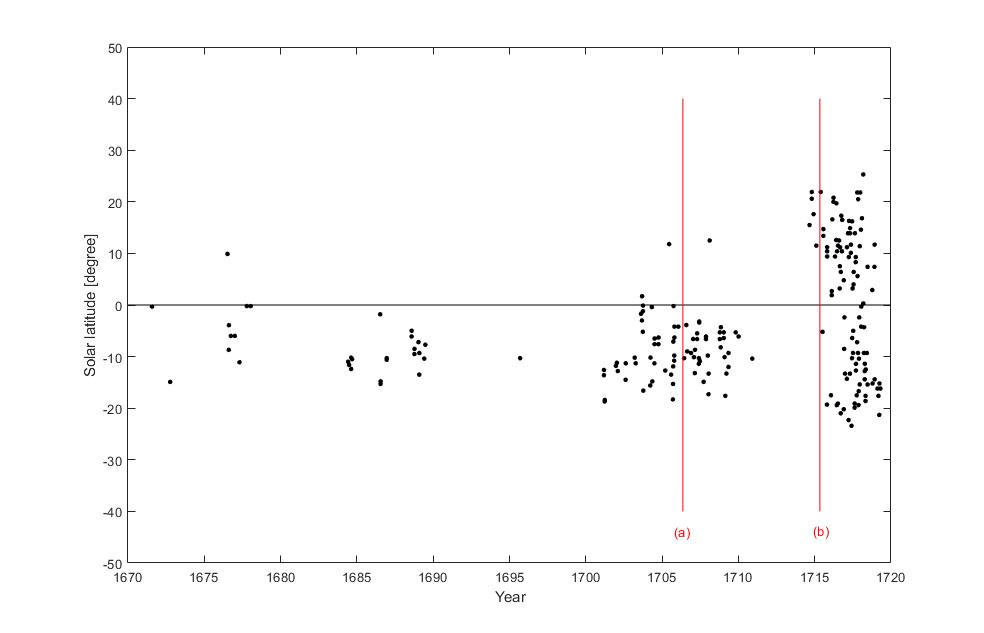}
    \caption{Sunspot positions in the late Maunder Minimum based on the data of Paris Observatory \citep{ribes1993} in comparison with the two cluster of graphical eclipse records in (a) 1706 (without significant visual streamers) and (b) 1715 (with significant visual streamers), as reproduced from \citet{hayakawa2021_eclipses}. }     
    \label{fig:solarlatitude}
\end{figure}

The transitions into and out of the Maunder Minimum have been confirmed to have been rather gradual (Fig.~\ref{fig:MM_SN_for_Jason_v2}) on the basis of contemporaneous sunspot records \citep{vaquero2011,hayakawa2021_JCM,Hayakawa2022_early18th_century}. Meanwhile, these transition periods are strikingly contrasted with the Maunder Minimum itself in terms of the presence of significant solar cycles, reported sunspot groups in both solar hemispheres \citep{Vaquero2015,munoz2019,Carrasco2019,hayakawa2021_eimmart}, and evident solar coronal streamers \citep{Hayakawa2020_Dalton_eclipse,hayakawa2021_eclipses}, as shown in Fig.~\ref{fig:solarlatitude}.

Overall, the solar activity level in the Maunder Minimum was extremely low (Fig.~\ref{fig:MM_SN_for_Jason_v2}).  The sunspot number estimates show the sunspot number between 1700 and 1726 substantially lower than the standard yearly value of the ISN V2 in later years \citep{clette2016}. This trend is consistent with the eclipse records \citep{Riley2015,hayakawa2021_eclipses} and the radiocarbon-based reconstructions \citep{Usoskin2021}. As such, the Maunder Minimum has been confirmed to be truly unique in the historical sunspot record, and has been labeled as a ``grand minimum'' to contrast it with other extended minima within normal solar cycles \citep{Usoskin2015,usoskin2023}.

\subsection{Modern Understandings from Proxy Data}

The Sun's global magnetic field in interplanetary space interacts with the Earth's magnetosphere in a way that modulates the arrival of charged cosmic ray particles.  These cosmic rays can strike atmospheric oxygen and nitrogen nuclei, shattering them and producing \ce{^{10}Be} via spallation. The record of historical atmospheric \ce{^{10}Be} variations can be calibrated against modern records to produce an additional proxy of the Sun's magnetic field.

Cosmogenic isotopes such as \ce{^{14}C} data from tree rings and \ce{^{10}Be} and \ce{^{36}Cl} data from ice cores confirm there was enhanced cosmic-ray flux during the Maunder Minimum, and thus lower solar activity \citep{Eddy76,Usoskin2015,Usoskin2021,Muscheler2007,Muscheler2016}. These cosmogenic isotope data also detected solar cycles during the Maunder Minimum \citep{beer1998}. Their general cycle duration had been off from regular 11-year cyclicity, either extended to $\approx$ 14 years on the basis of the tree-ring data \citep{miyahara2004} or shortened to 8-9 years on the basis of the sunspot records from telescopic observations, radio carbon data from another set of tree rings, and candidate auroral records indicated shorter solar cycles in the Maunder Minimum \citep{Vaquero2015,Usoskin2021,yan2023}. Further studies are needed to resolve this problem.

Is the Maunder Minimum unique? The Maunder Minimum is certainly unique in the historical sunspot record: even the Dalton Minimum or the deep cycle minima of Solar Cycles 23/24 and Solar Cycles 24/25 were not comparable to the depth of the Maunder Minimum \citep{hayakawa2021_eclipses,Hayakawa2020_Dalton_eclipse,Carrasco2022}. But what about on longer periods?

We have a little over 4 centuries of instrumental data \citep{Vaquero2009,Arlt2020,Clette2023,Hayakawa2024_Kepler} for a dynamo with a characteristic cycling time of 11 years, so not even 40 periods. Fortunately,  since the 1960s \citep{Stuiver1961,Lin1975, Stuiver1980} studies of cosmogenic isotopes have extended our chronological coverage to millennial timescales, and identified what appear to be similar events \citep{Stuiver1989,stuiver1991,voss1996}. Proxies like \ce{^{10}Be} and \ce{^{14}C} can trace the Sun's behavior through longer times at annual time resolution: roughly 8 centuries for the \ce{^{10}Be} data and 3 millennia for the \ce{^{14}C} data, at the time of writing \citep{beer1990,Muscheler2007,mccracken2015,Beer2018,brehm2021,brehm2025,Usoskin2021,usoskin2025}. In fact, analyses on the cosmogenic isotope data have revealed a number of grand minima in the past such as Sp\"orer Minimum, Wolf Minimum, and Oort Minimum \citep{Eddy76,Usoskin2007,Inceoglu2015,mccracken2015,Muscheler2007,Muscheler2016,silverman2021,usoskin2023}.

Physics-based models have dramatically improved such reconstructions, quantified a depth and a length of the grand minima in the Holocene, and enabled statistical studies on the grand minima in our Sun on millennial timescales \citep{Usoskin2007,Usoskin2014,Inceoglu2015,Wu2018}. These studies have listed up some 25 grand minima in the last eleven millennia as shown in Table 3 of \citet{usoskin2023}. These studies have tallied the time spent in grand minima to $\approx$ 1900 years within the last eleven millennia, implying that the Sun spent  $\approx$ 17 percent of the time in grand minima. \citet{Wagner2001} identified the 205 year de Vries cycle in \ce{^{10}Be} ice core data, which might indicate grand minima occur periodically on that timescale (although the Maunder minimum did not repeat 205 years later).

Among these periods, the last three millennia have been studied at annual resolution \citep{Usoskin2021,usoskin2025,usoskin2026} on the basis of precise radiocarbon measurements \citep{brehm2021,brehm2025,wang2026}. They have resolved some 270 solar cycles between 991 BCE and 1900 CE and detailed seven grand minima in this interval \citep{Usoskin2021,usoskin2025,usoskin2026}. Reconstructions for solar cycles detected with the highest confidence (having the highest quality flags) look slightly different from other radiocarbon-based reconstructions \citep{Miyahara2021} but are generally consistent with historical records for the coronal structures in the past eclipse records and the sunspot positions of camera-obscura-based pre-telescopic records \citep{Hayakawa2024_Kepler,hayakawa2024_early_MM,Hayakawa2025_eclipse}. These reconstructions assigned the lowest quality flags (mostly 0 and occasionally 1 out of 5) for the solar cycles during grand minima, where ``the cycle cannot be reliably identified'' (quality flag = 0) or ``the cycle is greatly distorted, at least one of its ends cannot be defined'' \citep{Usoskin2021}.
 
These data have allowed the scientific community to understand the grand minima not as products of the interplay of competing long-term modulations but as products of stochastic/chaotic processes. Their statistical distributions are inconsistent with those of the normal solar cycles, and the Sun's dynamo activity seems to have been in a special state in contrast with the normal solar cycles \citep{usoskin2026}. At the time of writing, \citet{usoskin2023} provides  the best reference for the reconstructions and interpretations of the grand minima based on the cosmogenic isotope data. 

\section{Dynamos}
 
One might interpret the Maunder Minimum as representing the breakdown of the solar dynamo and its cycles, and thus predict that Sun's magnetic field during this period was extremely weak and steady. Sunspot records indicate the presence of extremely small solar cycles \citep{Kovaltsov2004,Vaquero2015,Carrasco2022,hayakawa2024_early_MM,hayakawa2025_review}. In this case, the solar dynamo might have been still under operation, in a somewhat unique dynamo mode, but  \citet{Beer2018} showed that \ce{^{10}Be} records clearly show that the 11-year cycle was still active during the Maunder Minimum.

As we have seen, activity variations in the Sun can be at least partially reconstructed for the last $\sim 10^4$ years. In contrast, observations of the variable activity in other sun-like stars cover a few decades or less, but they represent a broad range in stellar parameters and ages. Two aspects of the underlying dynamo mechanism have key importance to forge a link between these distinct sets of observations. On the one hand, the response of dynamo models to changes in the control parameters can help explain the variety seen among stars. On the other hand, as the Sun's example shows, activity on one and the same star can take quite different characteristics when only observed for a shorter time, depending on the phase of its long-term activity variations. In particular, an understanding of the mechanism underlying grand minima, their occurrence rate and frequency would help place our Sun in the context of other stars.

\subsection{The effect of control parameters on the dynamo}

The solar dynamo is not fully understood, but a central feature of all models
for it is the generation of the Sun's magnetic field via the interaction of the
rotational and convective motions of its envelope \citep{charbonneau2020}.
Differential rotation winds up a weaker poloidal field into a strong toroidal
field. Loops formed on this field by flow advection or a buoyancy-driven
instability are then rotated out of the azimuthal plane due to the Coriolis
force ($\alpha$-effect), giving rise to an opposite polarity poloidal field,
closing the loop. The competition of this process of field generation with decay
due to turbulent magnetic diffusion is characterized by the dynamo number
$D=\alpha\Omega R^3/\eta^2$ ($R$ is the stellar radius, $\Omega$ is the angular frequency, and the parameter $\alpha$ has units of velocity). The turbulent
diffusivity may be estimated as $\eta\sim H^2/\tau_c$ where $H$ is the scale
height and $\tau_c$ is the convective turnover time in the deep convective
zone. As $\alpha$ is also determined by rotation, dynamos of
solar type stars ultimately differ in their mass and rotational velocity only.
The mass $\mass$ determines all other parameters ($R$, $H$, $\tau_c$, and the
thickness of the convective shell) via the stellar model.  The two time scales
$\tau_c$ and period $P=2\pi/\Omega$ will only enter the problem via their ratio, the
Rossby number Ro$=P/\tau_C$. 

The large-scale flows (differential rotation and meridional flow) are then
determined by the underlying parameters $\mass$ and Ro. Recent numerical
simulations of rotating convective envelopes \citep{Brun2017} have led to major
advances in our understanding of this process, showing the existence of three
well-defined regimes in Ro. As magnetic braking leads to the gradual increase of
Ro, these regimes may also be thought of as corresponding to stages of a
temporal evolution of the flow pattern in a star's lifetime.  In the fast
rotating regime the flow pattern displays several cells in the meridional plane,
with alternating faster and slower rotating belts, reminiscent of what is seen
in gas giants planets like Jupiter. At moderate rotation rates a solar-like differential rotation
ensues, with a single-celled meridional circulation. In the slowest rotating
regime antisolar differential rotation is seen in hydrodynamical models.

Building upon these results, \cite{Noraz2024} recently extended the simulations to
the magnetic case, studying the behaviour of the corresponding stellar dynamos.
Fast rotators were found to display short cycles with butterfly wings of
different cycles coexisting at different latitudes. For moderate rotation
solar-type dynamo behaviour was qualitatively reproduced, while for Ro exceeding
the solar value cyclic behaviour disappears and a steady activity ensues.  

Whether or not the antisolar differential rotation with a steady level of
activity can be realized is currently a topic of active research. The 
solar--antisolar transition implies a vanishing $\omega$-effect, reducing  the
dynamo number below critical, so magnetic braking would not continue. The
observations pointing to weakened magnetic braking for Ro$/$Ro$_\odot\!\!\sim$1
provide support for this scenario, further suggesting that the Sun is near the
final stage of being able to sustain a dynamo \citep{Metcalfe2025b}. This would seem to imply that the
antisolar regime is never reached. Nevertheless, some studies
\citep{Brandenburg2018, Mathur2025} point to an {\it increase} in the level of
activity in stars with Ro exceeding the solar value, potentially explained by
the development of a dynamo driven by antisolar differential rotation. Reaching
this regime may be facilitated by slow structural changes related to stellar
evolution, especially during subgiant expansion \citep{Metcalfe2026}.  On the
other hand, the recent state-of-the-art MHD simulations of \cite{Hotta2026} do
not display an antisolar profile for any value of Ro, further deepening the
puzzle.

The role of the mass parameter $\mass$ in these simulations is limited: in fact,
even the threshold values in Ro where transitions between these domains take
place do not show a clear dependence on $\mass$ \citep{Noraz2024}. On the other
hand, there are observational indications that the level of activity, the rate
of magnetic braking and even the cycle period do seem to depend on metallicity
\citep{Karoff2018,Carvalho2025}.

However suggestive these simulations are, it must be added that they are still
subject to serious limitations. They are still very far removed from actual
stars in the parameter space: they are much less stratified and much more
diffusive, hence intermittent structures are much less well resolved and
non-local effects are more limited. An outstanding unresolved problem is the
{\it convective conundrum:} helioseismic constraints indicate that turbulent
flow velocities in the deep convective zone are significantly lower than in
simulations \citep{OMara2016}. The simulations described above can, in fact,
only be aligned with observational constraints by arbitrarily rescaling Ro to
account for this unresolved issue.

Mean-field models of the internal flow pattern \citep{Kitchatinov2011} have been able to qualitatively reproduce the solar-type regime corresponding to moderate rotation. However, significant quantitative discrepancies remain and the the other two regimes have not been reproduced. Nevertheless, stellar mean-field dynamo models based on this approach \citep{Vashishth2023} have been able to reproduce some observed trends in stellar dynamos.

In solar dynamo models the differential rotation profile is normally prescribed, implying that these models can only be used to represent other solar-type stars in a relatively narrow parameter range. Within this range the changing behavior of the models with dynamo number may nevertheless be considered as representative of the evolution of the dynamo. We introduce $\delta=D/D_{\mbox{\scriptsize cr}}-1$ as a measure of the supercriticality of the dynamo. ($D_{\mbox{\scriptsize cr}}$ is the critical dynamo number.) 
For relatively faster rotation the dynamo will be in the supercritical domain $D\gg 1$; later, with magnetic braking it will reach the near-critical regime $\delta\la 1$. In current thinking of the solar dynamo one hotly debated issue is in fact its current degree of supercriticality \citep{Karak2023}. 

\subsection{Supercritical dynamos: supermodulation due to multiple unstable modes}

For increasing dynamo numbers, a series of bifurcations leads to the coexistence
of several unstable oscillatory solutions. These solutions typically differ in
their geometrical structure: the shorter geometrical scales of higher order
multipoles imply more effective diffusive damping and consequently a higher
critical dynamo number required for their growth. Hence, the two lowest order
modes, dipole and quadrupole, will be normally the first ones to appear. In the
nonlinear domain some parts of the parameter regime in many such models can
display chaotic behavior, spontaneously transitioning between states resembling
the linear modes. Indeed, it was first suggested by \cite{Tobias1997} and
\cite{beer1998} that this may give rise to large-amplitude variations of the
solar cycle on a longer time scale. Such variations manifest in occasional grand
minima  which, in this scenario, correspond to transitional states between
dipole and quadrupole solutions, naturally explaining the hemispheric asymmetry
of the low-level activity  during the Maunder Minimum. This idea was later
further elaborated \citep{Weiss2016,Beer2018} to explain the long-term periodic
and aperiodic supermodulation seen in the $\sim 10^4$-year cosmogenic record.

Another similar mechanism of supermodulation has been suggested in the context of a strongly supercritical flux transport dynamo by \cite{Thibeault2023}. Here the appearance of further bifurcations is related to the time delay inherent to these models, and transition between the coexisting modes is ascribed to stochastic forcing.

\subsection{Near-critical dynamos: stochastic driving of grand minimum episodes}

In the turbulent convective zone stochastic fluctuations in model parameters are naturally expected. On observational and theoretical grounds the strongest fluctuations are expected in $\alpha$ and in the meridional flow speed. In flux transport dynamos fluctuations in $\alpha$ are manifest in the large scatter around the observed latitude--tilt relation of solar active regions (Joy's law). Flux transport dynamos and indeed even simple surface flux transport models have had impressive success in quantitively reproducing the observed variations in the sunspot cycle record \citep{Bhowmik2023,Cameron2023,Yeates2023}. It has been found that, due to the nonlinear character of the Babcock--Leighton $\alpha$-mechanism in such models, such stochastic fluctuations can in extreme cases even lead to the total supression of the dynamo \citep{Nagy2017}. The nonlinear nature of the $\alpha$-effect constitutes a challenge in explaining the recovery of the dynamo after such grand minimum events without external ``help'', e.g. from a linear $\alpha$-effect assumed to operate on weak fields in the deep convective zone \citep{Hazra2014}. However, in an important study \cite{KarakMiesch2018} demonstrated the possibility for a flux transport dynamo to recover from stochastically induced grand minima if downward directed pumping of the magnetic field is taken into account. In such a situation the decay of the poloidal field component is suppressed, so it keeps producing active regions at a low level until appropriate fluctuations allow the system to reach criticality again. This model also convincingly reproduces observed features of the grand minima such as low-level cyclic activity and hemispheric asymmetry.

\subsection{How supercritical is the solar dynamo?}

Which of the above possibilities is realised in the Sun? As mentioned above, the low-level cyclic activity and the hemispheric asymmetry suggested by observations during the Maunder Minimum can be reproduced in both classes of models, so these observations are not decisive. 

One argument in favour of the supercritical case is the apparent presence of
numerous secular periodicities in the long-term solar activtiy, reconstructed
from the cosmogenic radionuclide record. As described in Section 2, the power spectrum of the record shows
peaks around 88 and 205 years, among others. As in the near-critical case grand
minima are stochastically forced, such secular periodicities are not expected to
be present there. Note, however, that the apparent sharpness of these peaks on a
linear frequency scale (e.g. Fig.~10 of \cite{McCracken2013}) is misleading. In a
more sobering logarithmic representation, as in Fig.~22 of \cite{usoskin2023} these peaks are much broader and they barely reach the significance limit. Indeed, in a generic normal form model for a stochastically forced, weakly nonlinear limit cycle \cite{Cameron2017} find that in a time series of length comparable to the cosmogenic record peaks in a number and at a level comparable to the actual record readily appear. 

The degree of supercriticality may be tested by the rate of recovery from grand
minima. With this method, \cite{Wavhal2025} find that the solar dynamo is two
times critical, i.e.\ $\delta\simeq 1$. The threshold value of $D$ for the
second bifurcation varies rather widely among dynamo models but it is normally
well above this value. E.g., in \cite{Thibeault2023} multiple unstable modes
typically appear for $\delta\ga 5$ except for bulk diffusivities $\eta <
10\,$km$^2/$s in the convective zone, a likely unrealistically low value. Thus,
while we admit that this issue is not settled, in view of the above
considerations and of the success of the \cite{KarakMiesch2018} model we will proceed
under the working hypothesis that the Sun's dynamo is near-critical and grand
minima originate from stochastic perturbations. It is in this framework that we
will suggest in Section 5 a scenario for how stellar dynamos grow gradually
``frustrated'' as they approach criticality.

\section{Stellar Magnetic Fields and Cycles}

We would learn much from a more extended record of solar sunspot number. Fortunately, we have another proxy that not just extends a record of solar-like cycles in time, but shows us how it might vary with other properties of the Sun, and on timescales of \textit{billions} of years: chromospheric activity records of Sun-like stars.

We have learned much about solar and stellar dynamos from sun-like stars. While starspots are very challenging to detect on stars as old and inactive as the Sun, especially over decade timescales, we can measure and observe variations in the strength of emission cores from the \CaHK lines at 393 and 397 nm \citep{Wilson1968}, which are generated by the cooling of stellar chromospheres. Because this cooling must be, in the long run, balanced by the heating of these chromospheres, and because the heating is caused by magnetic fields \citep{Leighton1959}, these records constitute a very similar proxy to sunspot number for the global surface magnetic field strength of the star.

Another aspect of magnetic field strength that affects observable properties of stars is related to stellar winds. The mass lost by stars to their winds is ionized and so trapped on the star's magnetic field lines, which spiral out from the stellar surface due to star's rotation. Conceptually (but not precisely), we can think of these lines as co-rotating with the star out to the Alfv\'en radius, and so these particles gain angular momentum as they move away from the star and slowly tap the star's spin energy \citep{WeberDavis1967, Kawaler1988}.  The result is that Sun-like stars spin down from rotation periods of a few days to several weeks over their Main Sequence lifetimes \citep[\eg][]{MacGregor1991}. The goal of gyrochronology \citep[\eg][]{Barnes2007,Barnes2010} is to determine the ages of stars from these spin rates, which is otherwise an extremely challenging task \citep[see][]{Soderblom2010}.

Importantly, this spindown mechanism only operates on cool stars (roughly surface temperatures between 3400--6700 K) which have convective envelopes and radiative cores \citep{Kraft1967}, in part because the magnetic dynamo of these stars is governed by the dynamics at the base of the convective zone, and because the transfer of angular momentum from the interior of a star depends on these details of its internal structure. Very cool stars (which are fully convective) appear to be less efficient at spinning down, and hotter stars with radiative envelopes generally barely spin down at all before exhausting their central hydrogen and evolving off the main sequence.

\citet{Skumanich1972} pioneered gyrochronology by identifying what appeared to be a square-root dependence of rotation period on age ($P_\mathrm{rot}\propto\sqrt{\mathrm{age}}$) in Sun-like stars.  Today, studies of star clusters with well-known ages have allowed for more detailed studies that have revealed important nuances to this law. The spindown of young stars, especially, has proven to be strongly mass-dependent \citep{Bouvier2008}.   

Studies of stellar activity, rotation, and age have confirmed the central role of the Rossby number in stellar dynamos \citep{DurneyLatour1978}.  It is now well established that young stars spin rapidly and so have powerful dynamos that generate strong surface magnetic fields, and that these fields weaken as the star slowly spins down over billions of years \citep[e.g.,][]{See2019}. 

Eventually, these stars begin to exhaust their central hydrogen supply, their luminosity increases by tens of percent, causing their radii to increase, first very slowly, but eventually, as the central hydrogen concentration goes to zero, more rapidly. This is the transition to the relatively short-lived subgiant phase, during which stellar cores switch to hydrogen shell burning and they swell to become red giants. Today, distance measurements are precise enough that subgiant stars can be unambiguously identified by their higher luminosities \citep{GaiaDR3}. Such subgiants generally rotate slowly and so have less intense dynamos, rarely show significant magnetic activity at all, and exhibit some of the lowest chromospheric emission levels of any stars (\citealt{Wright04b}, but see \citealt{Metcalfe2020, Metcalfe2024b} for exceptions). Indeed, the magnetic states of subgiant stars are reminiscent of the Maunder Minimum.

\section{Chromospheric Activity Monitoring of Sun-Like Stars}

\subsection{The Mount Wilson and Other Surveys}

An important and pioneering work in understanding the stellar context of the Sun's magnetic field strength is the Mount Wilson \CaHK survey started by Olin Wilson \citep{Wilson1968} and carried on by others including \citet{Vaughan1978}, \citet{Duncan91}, and \citet{Baliunas95b}.  Over the course of 35 years this spectroscopic survey collected a continuous stream of data on the strength of the \CaHK lines in around 100 stars, establishing that decade-long magnetic cycles were a ubiquitous feature of middle-aged, Sun-like stars.  

Such measurements are expressed as the strength of the \CaHK lines in one of two forms, either the $S$ index, which is an arbitrarily calibrated dimensionless quantity, or \rphk, a representation of the fraction of a star's total luminosity being emitted in the \CaHK lines \citep{Noyes1984}.

Solar and stellar magnetic cycles are also apparent in the total luminosity, measured for the Sun as variations in the total solar irrandiance (TSI) \citep{kopp2025} and in stars by long-term photometry. Together, records of photometry and \CaHK measurements have helped generate a comprehensive picture of how Sun-like star's magnetic activity levels vary with time \citep[e.g.][]{Henry1995_ApJS,Henry95,Radick2018}.

The discovery of exoplanets around Sun-like stars in 1995 ushered in a new era of intense spectroscopic monitoring of such stars to search for the subtle Doppler shifts that reveal exoplanets' gravitational pull on their host stars \citep{Mayor_queloz}. An ancillary benefit of such monitoring was that many planet-finding spectrographs included the \CaHK lines in their bandpass, allowing all planet-search target stars' magnetic field strengths to be monitored \citep[both for the interest of stellar astrophysicists but also because the Doppler noise or ``jitter'' in these stars depends on stellar surface magnetic fields, \eg][]{Wright05}.  

When the Mount Wilson survey ended around 2004, some programs picked up many of its target stars to ensure their time series would not be truncated.  For instance, \citet{Wright04} measured and curated the \CaHK line strengths in all of the \about 18,000 spectra collected to date from over 1,000 stars as part of the California and Carnegie Planet Search Program, a project that has been continued by \citet{Isaacson10} and others.  Some of the other programs conducting similar work are  \citet{Henry96}, \citet{Strassmeier00}, \citet{Tinney02}, \citet{Hall2007}, \citet{Schmitt2014}, \citet{GomesdaSilva2021},and \citet{Morris2025}.  

The various activity monitoring programs occasionally suffer from subtle calibration mismatches and data errors that make direct comparison and interpretation challenging. \citet{Baum2022} curated and harmonized the activity records of the Mount Wilson survey and two different calibrations of the California Planet Survey to produce a unified activity history of 59 Sun-like stars with typical timespans of 50 years.

The sample of stars with good activity histories is heterogeneous, and so its sample biases are difficult to quantify. The original Mount Wilson survey sought the brightest ``Sun-like" stars visible from Southern California, limited at the hot end by the presence of narrow lines and significant \CaHK\ emission (roughly spectral type F2) and on the cool end by apparent magnitude \citep{Wilson1978}.  Efforts were also made to include a balance of strong- and weak-emission line stars, which resulted in an overabundance of young stars in the sample.  Precise distances were not available when the Mount Wilson survey entered its various phases, and as a result it inadvertently included a substantial number of subgiant stars. 

The planet-search programs overwhelmingly focused on older, slowly rotating, inactive stars because they exhibit the lowest levels of stellar ``jitter'' and allow for the most precise radial velocity measurement.  These selection biases make it challenging to perform robust statistical tests or precise conclusions from stellar activity records.  Nonetheless, the richness of the data set does allow for a rough universal model of stellar activity variations for Sun-like stars, and for the identification of particularly revealing case studies.

\subsection{The Search for Maunder Minimum Analog Events}


Because many subgiants had been inadvertently included in early activity monitoring programs, multiple surveys found a significant minority of nearby ostensibly Sun-like stars to be in very low activity states reminiscent of the Maunder Minimum.  For instance \citet{Baliunas90} found 30\% of stars to be in such a state, \citet{Saar92} found the number to be 10--15\%, and other contemporary surveys found similar numbers \citep{Henry96,Baliunas95b, Saar98,Gray03}. The implication of these surveys was that the Sun likely spends roughly this fraction of its time in such a state.  Much work also went into studying how these stars' properties might be different from their cycling counterparts, with implications for the Sun's effect on Earth climate and space climate \citep[see, \eg][ and references therein.]{Basri2013,Reinhold2020,Connolly2021,Nandy2021,Reinhold2021}

\citet{Wright04b} compiled newly-measured precise parallaxes from the ESA Hipparcos mission \citep{PerrymanESA} for the stars in these catalogs and found that, nearly without exception, all of the so-called Maunder Minimum stars were significantly evolved and transitioning into a subgiant state.  That is, their low and steady levels of chromospheric activity were characteristic of very old, slowly spinning stars, and so not good analogs to the sort of grand activity minimum characterized by the Maunder Minimum in the current (middle-aged) Sun.

This complicated searches for Maunder Minimum analog events in Sun-like stars because it meant that simply identifying a ``flat activity'' or low-activity star was insufficient.  \citet{SaarTesta2012} suggested searching for stars that 1) were bona fide dwarf stars  2) with activity levels lower than similar cycling stars' minima and 3) had very low activity level variation.  

\citet{Shah2018} took a different approach, following \citet{Donahue1995} who had argued that the star HD 3651 might be entering a Maunder Minimum phase based on a decrease in cycling activity seen in the Mount Wilson sample.  \citet{Shah2018} used California Planet Survey data to show that, while that star's cycle amplitude had decreased significantly in the 2000's, it was still clearly cycling. Shah established that activity records for Mount Wilson legacy stars were then long enough that a star might be seen ceasing (or starting) cycling behavior, and suggested HD 4915 as a candidate \citep[which][would show has continued to cycle]{FloresTrivigno2024}. 

\subsection{HD 166620: A Star in a True Magnetic Grand Minimum}

\begin{figure}
    \centering
    \includegraphics[width=1\linewidth]{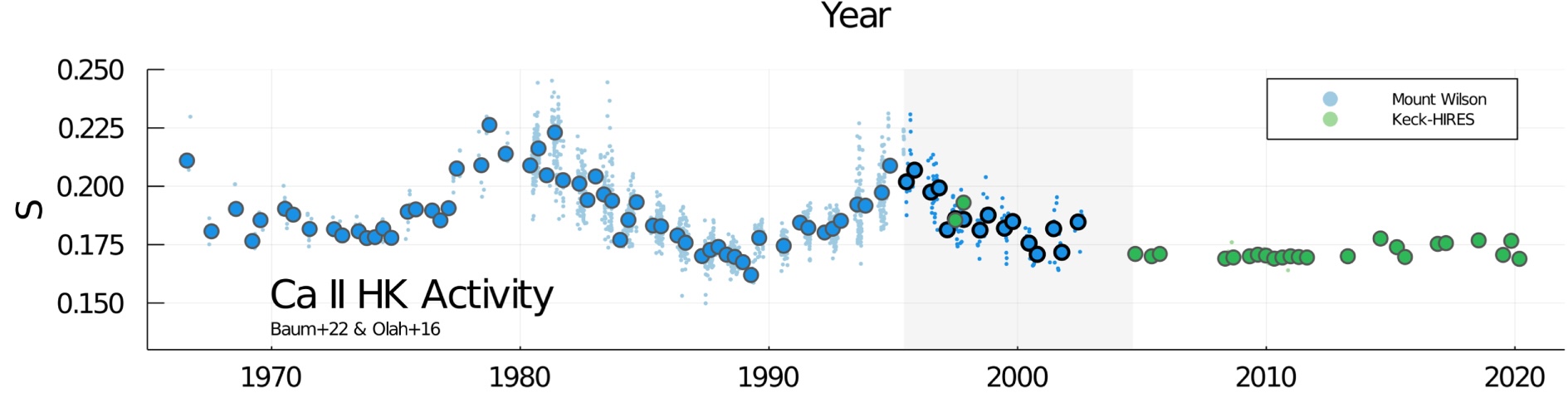}
    \caption{Panel 2 of Fig.~1 from \citet{Luhn2022} showing the combined chromospheric activity record (as the $S$ index) of HD 166620, using data from \citet{Baum2022} and \citet{Olah2016}.}
    \label{fig:LuhnFig1}
\end{figure}

The 50+ year compilation of activity records by \citet{Baum2022} included the star HD 166620, which they showed had apparently completely and dramatically changed its behavior around 2004 from strongly cycling to completely flat. \citet{Luhn2022} identified additional chromospheric activity and photometric records that firmly established the star's transition to a flat-activity state, making it probably the first and only clear example of a star in a Maunder Minimum-like grand minimum (see Fig.~\ref{fig:LuhnFig1}).

Interestingly, the activity level in HD 166620 is not significantly lower than its last cycle minimum in 1989 (although it is lower than the previous cycle minimum in 1974).  

HD 166620 is not a perfect solar analog.  It is cooler and less massive than the Sun ($T_\mathrm{eff}=4970$ K, spectral type K2{\sc V}), and has slightly sub-solar metalicity ([Fe/H] = $-$0.16) \citep{Baum2022}.  Its rotation period is between 42.4 days \citep{Donahue1996} and 45.06 days \citep{Luhn2022}, which would make it older than the Sun. 

Despite these differences, \citeauthor{Luhn2022} note that these stellar parameters imply a Rossby number very similar to that of the Sun, implying it has a similar dynamo state. Multiple campaigns to monitor HD 166620 are now underway from the ground and space \citep{Bennedik2026, Wood2026, Chiti2026}, both to establish its chromospheric activity behavior during this minimum and to see when it begins cycling again, but also to understand its overall magnetic state.  Fortunately, such a quiet star makes an excellent target for precise radial velocity work to find exoplanets, so the star is sure to receive plenty of attention from high resolution spectrographs for many years to come.

\section{Lessons From the Stars}

\subsection{Magnetic Grand Minima}

It can be challenging to compare subtleties in activity measurements across instruments, but \citet{Luhn2022} argued that the lesson of HD 166620, presuming its entry to a grand minimum is a good analog of the Maunder Minimum, is that the transition from a cycling to a grand minimum state can be relatively abrupt, and that while the minimum state has less variability than a typical minimum, its overall level is not extraordinarily low.  These data points can presumably inform dynamo models that attempt to explain grand minima events in the Sun \citep{Weiss2016}.

One property of grand minima that the data set of \citet{Baum2022} can address is their duty cycle---that is, how often might we expect the Sun to enter into such a state? This is probably best constrained by terrestrial records, but the combined time series for the stars offers a path to check this number across a range of Sun-like stars.

Of the 59 stars with lengthy records studied by \citet{Baum2022}, each with around 50 years of data, only one is clearly shown entering (or leaving) a Maunder-minimum like state. This implies it is quite a rare event, happening around once every 3,000 years.  

Nonetheless this number stands in contrast to the 205 year de Vries cycle and the average period of 440 years between minima listed by \citet{usoskin2023}. The two explanations are likely some combination of the Maunder minimum being an extraordinary example of a grand minimum, occurring more rarely than the grand minima traced by cosmogenic records, and the stellar data set containing stars with a range of dynamo characteristics, most of which do not experience Maunder minimum-like events.

\subsection{Variability in the Cycling Behavior}

More broadly, dynamo models can be tested against the behavior of other single, Sun-like stars with slightly different masses, temperatures, metallicities, ages, and rotation periods.  The first such studies were that of \citet{Wilson1978} and \citet{Baliunas95b}, who established that younger stars (ages of $10^7$--$10^9$ years) exhibit high chromospheric activity variability that is generally stochastic, showing no cycles. Intermediate-age dwarf stars (1--2 Gyr) have characteristically lower levels of activity and short, occasional cycles, while older dwarf stars (like the Sun) have the lowest activity levels and characteristically smooth, decade-long cycles \citep{Noraz2024}.

More recently, \citet{Isaacson2024,Isaacson2025} analyzed the cycling behavior of the stars in the California Legacy Survey (an extension of the California and Carnegie Planet Search Program). One of the strongest trends they find is that cycle amplitude in their sample is well determined by the surface (effective) temperature of stars. This parameter most strongly traces stellar mass in dwarf stars, but is also sensitive to age in stars like the Sun. \citet{Isaacson2024} found that lower temperatures were associated with larger amplitudes, and that for stars of a given mass cycle amplitude generally decreases slightly with age.

They find that the cycle periods for more active stars are shorter (typically between 5--10 years) and that period generically increases as stars become more inactive \citep[see also][]{Metcalfe2026b}. For lower-activity stars (which are presumably older) there is a wide variation in cycle period, much of which is a function of stellar mass: lower-mass stars show longer periods (around 20 years) and stars more massive than the Sun show periods as short as 5 years.  But generally they find that correlations between cycle period, cycle amplitude, mean activity level, and activity variations with stellar properties to be rather weak.  

The \textit{Kepler} mission performed precise photometric monitoring for over 100,000 stars and was able to measure rotation periods for many of them \citep{McQuillan2014}, providing novel insights into the functional form of stellar spindown at intermediate stellar ages.  It found that the magnetic braking, too, begins to fail near the critical Rossby number of around 0.9 in Solar units \citep{vanSaders2016, Metcalfe2025b}, causing stars to stall at a particular rotation period that depends on the depth of their outer convective zones (and thus their mass).  In some stars, this precise photometry permits the measurement of asteroseismic modes so that a star's convective zone depth and age can be independently determined \citep{Metcalfe2014, Creevey2017}.

\subsection{An Emerging Model}

\begin{figure}[ht!]
    \centering
    \includegraphics[width=1\linewidth]{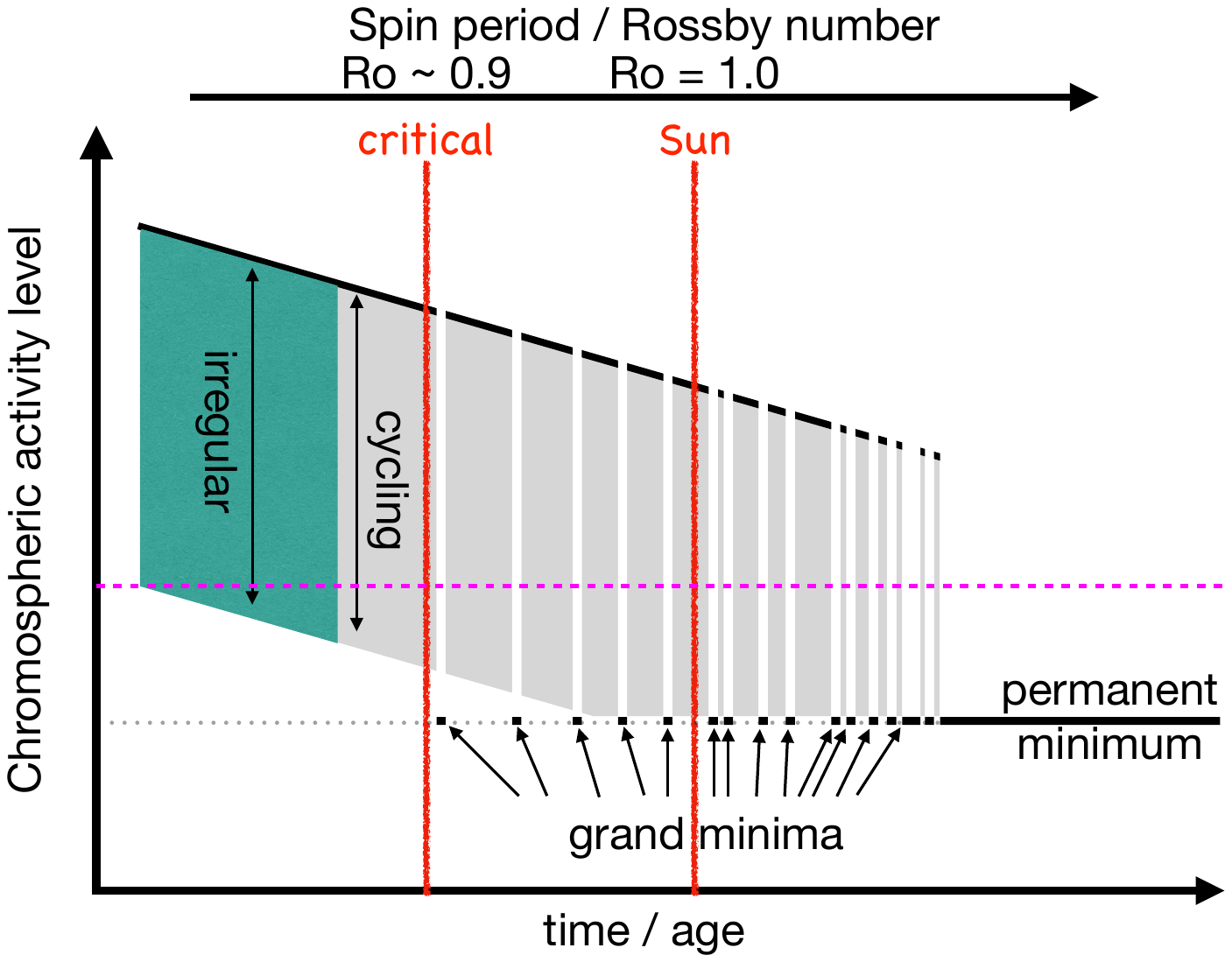}
    \caption{Schematic showing an emerging model for the evolution of cycling behavior in Sun-like stars. Stars' chromospheric magnetic activity level drops with time as their Rossby number \Ro increases over billions of years. During this time, their chromospheric activity levels cycle between slowly decreasing maxima and minima with periods of around a decade.  As the Rossby number exceeds a critical value of $\Ro=0.9$ (in units where the Sun has $\Ro=1$), occasional periods of frustration appear when the star's cycle disappears and its surface magnetic activity level is constant at a grand minimum, akin to the solar Maunder Minimum (recent results from \citet{Metcalfe2025b} suggest this critical value could be slightly higher than the solar value). As the star continues to age these episodes become more frequent and the cycle activity level minima reach the grand minimum level and stay there, until at some point cycling behavior ceases entirely and the star enters a permanent activity minimum. The mean activity level of the Sun, near $\Ro = 1$ is shown as a horizontal magenta dashed line.
    This diagram is purely schematic. To point out just a few simplifications: cycle amplitudes are variable on many timescales; the number of grand minima is vastly larger than shown here; and the magnitude and functional form of the overall decrease in activity levels is more complex than shown.}
    \label{fig:sputter}
\end{figure}

Despite their uncertain selection biases, studies of stellar activity levels reveal an emerging picture---still fuzzy but coming into focus---of stars that begin their lives rapidly rotating and magnetically active with irregular variations, and steadily spin down and settle into regular cycling behavior \citep{Egeland2017proc}. Upon reaching a critical rotation period as a fraction of its convective turnover time, the surface magnetic cycle begins to sputter and exhibit decades-long grand minima \citep{Vashishth2023}.  As the star continues to spin down, these episodes become more frequent, until eventually the starspot cycling behavior ends entirely \citep{Metcalfe2017, Metcalfe2025b}. Figure~\ref{fig:sputter} shows, very schematically, this emerging picture.

\section{Future Opportunities}

A key parameter that will help clarify this picture is precise ages, masses, and rotation periods for Sun-like stars with good activity records. For a long time, stellar clusters were by far the best laboratories for such work because the stars in them differ only in their masses and binarity, allowing their compositions and ages to be precisely determined. Clusters of intermediate-age stars like the Sun are rare, however, and generally too far away to permit close inspection. 

Stellar ages are notoriously hard to determine for field stars (\ie those not in clusters). In some cases, dense, precise photometry or radial velocity measurements can also detect the strongest asteroseismic modes in stars, providing an independent age constraint \citep{Bedding04,Kjeldsen05,Aerts2021}.  Rotation periods can be gleaned from \CaHK time series and precise photometry, the latter of which has been applied to stellar clusters to help provide a picture on the effects of age and composition on patterns of stellar spin-down \citep[\eg][]{Curtis2020,Carvalho-Silva2025AgeActivity}.   

Collectively, this work is bringing us closer to the goal of precise gyrochronology, which will greatly improve our ability to understand the selection biases of the stellar activity monitoring campaigns' samples, and to accurately place the Sun's age, rotation, and long-term magnetic behavior in the context of its peers. Eventually, studies of stellar and Solar activity through the dimensions of time, mass, and composition may provide a robust and precise model of stellar dynamos, and help us better understand the past, present, and future of stellar magnetic fields on the Earth and Earth-like planets throughout the Galaxy.

\begin{acknowledgements}
    This review made use of NASA's Astrophysics Data System Bibliographic Services. HH thanks V. M. S. Carrasco for providing his datasets.
\end{acknowledgements}

\begin{funding}
    T.S.M.\ acknowledges NASA grant 80NSSC25K7563.  K.P. acknowledges support from NKFIH grants TKP2021-NKTA-64 and 2026-4.1.1-MISSZIÓK-2026-00005. H.H. thanks financial supports of JSPS Grants-in-Aid JP25K17436 and JP25H00635, the ISEE director’s leadership fund for FYs 2021--2025, the Young Leader Cultivation (YLC) programme of Nagoya University, Tokai Pathways to Global Excellence (Nagoya University) of the Strategic Professional Development Program for Young Researchers (MEXT), the young researcher units for the advancement of new and undeveloped fields in Nagoya University Program for Research Enhancement, the Interdisciplinary Research Strategy Projects of the Institute for Space--Earth Environmental Research (ISEE), Transdisciplinary Network linking Space-Earth Environmental Science, History, and Archaeology (JPMXP1324134720) of MEXT Promotion of Development of a Joint Usage/Research System Project: Coalition of Universities for Research Excellence Program (CURE). 
\end{funding}

\begin{conflictofinterest}
    The authors declare no Conflict of Interest.
\end{conflictofinterest}

\begin{dataavailability}
All data in this review are previously published and available at the citation given.
\end{dataavailability}


\end{document}